\documentclass{eas}
\usepackage{graphicx}

\begin{document}

\title{Light propagation through large-scale inhomogeneities in the Universe and its impact on cosmological observations} 

\runningtitle{Bolejko: Cosmological observation in the inhomogeneous Universe}

\author{Krzysztof Bolejko}\address{Nicolaus Copernicus Astronomical Center, Polish Academy of Sciences, Bartycka 18, 00-716 Warsaw, Poland;
\email{bolejko@camk.edu.pl}}

\begin{abstract}
This paper analyses cosmological observations within inhomogeneous and exact solutions of the Einstein equations. In some way the analyses presented 
here can be freed from assumptions such as small amplitude of the density contrast.
The supernova observations are analysed using the Lema\^itre--Tolman model 
and the CMB observations are analysed using the quasispherical Szekeres model.
The results show that it is possible to fit the supernova data
without the cosmological constant. However if
inhomogeneities of sizes and amplitudes as observed in the local Universe
are considered, their impact on cosmological observations is small.
\end{abstract}

\maketitle

\section{Introduction}

The Universe which we observe is very inhomogeneous. Among the structures in the Universe exist groups and clusters of galaxies, large cosmic voids and very large elongated structures such as for instance filaments. In cosmology, however, the homogeneous and isotopic models of the Robertson-Walker class have been used almost exclusively.
Namely cosmological observations are usually analysed within the homogeneous framework.
If cosmic structures are taken into account they are usually described by an approximate perturbation theory. This works well as long as the perturbations remain small, but cannot be applied once perturbations become large and evolution becomes non-linear.  
That is why it is important to analyse cosmological observations not only in homogeneous models
but also inhomogeneous ones.

In this paper two types of  inhomogeneous and exact solution of the Einstein equations are considered. The spherical symmetric Lema\^itre--Tolman model will be employed to analyse supernova observations
and the quasispherical Szekeres model to analyse the Cosmic Microwave Background (CMB) radiation.

\section{The Szekeres and Lema\^itre--Tolman models}

The metric of the Szekeres model (Szekeres 1975) is of the following form:

\begin{equation}
ds^2 =  c^2 dt^2 - \frac{ \left(\Phi'(r,t) - \Phi(r,t) \frac{\textstyle E'(r,p,q)}{\textstyle E(r,p,q)} \right)^2}
{\varepsilon - k(r)} dr^2 - \Phi^2(r,t) \frac{(dp^2 + dq^2)}{E^2(r,p,q)}, \label{dshk}
 \end{equation}
where ${}' \equiv \partial/\partial r$, $\varepsilon = \pm1,0$ and $k \leq \varepsilon$ is an arbitrary function of $r$, $\varepsilon = 0, \pm 1$. The case when $\varepsilon = 1$ is often called the quasispherical
Szekeres model.

The Einstein equations reduce to the following two:

\begin{equation}
\kappa \rho c^2 = \frac{ 2 M' - 6 M E'/E}{\Phi^2 ( \Phi' - \Phi E'/E)},~~~~~
\frac{1}{c^2}\dot{\Phi}^2 = \frac{2M}{\Phi} - k + \frac{1}{3} \Lambda
\Phi^2,
\end{equation}
where $\dot{} \equiv \partial/\partial t$,  $\rho(r,t,p,q)$ is matter density, 
$M(r)$ is active gravitational mass, and
$\kappa = 8 \pi G / c^4$.

The quasispherical Szekeres model is the generalisation
of the Lema\^{\i}tre-Tolman (Lema\^itre 1933; Tolman 1934)
model and reduces to it when functions $E'=0$.
For detail description of these models the reader is refered to 
textbook by Pleba\'nski \& Krasi\'nski (2006).

\section{Light propagation}

Light propagates along null geodesics. If  $k^{\alpha}$ is a vector tangent to a
null geodesic, then: $k_{ \alpha ; \beta} k^{\beta} = 0$.
The redshift formula is:

\begin{equation}
1 + z = \frac{\left( k^{\alpha} u_{\alpha} \right)_e}{ \left( k^{\alpha}
u_{\alpha} \right)_o}, 
\end{equation}
where the subscripts $_e$ and $_o$ refer to instants of emission and
observation, respectively.
Since the temperature scales as (1+z),  $T_e/T_o = 1 + z$
the temperature fluctuations measured by comoving observer are:

\begin{equation}
\left( \frac{\Delta T}{T} \right)_o = \frac{1/(1+z) -
1/(1+\bar{z})}{1/(1+\bar{z})} + \left( \frac{\Delta T}{T} \right)_e
\frac{1+\bar{z}}{1+z}. 
\end{equation}
where quantities with bars $~ \bar{}~$ refer to the average quantities, i.e. the
quantities obtained in the homogeneous Friedmann model.

\section{Cosmological observation}

\begin{figure}
\begin{center}
\includegraphics[scale=0.4]{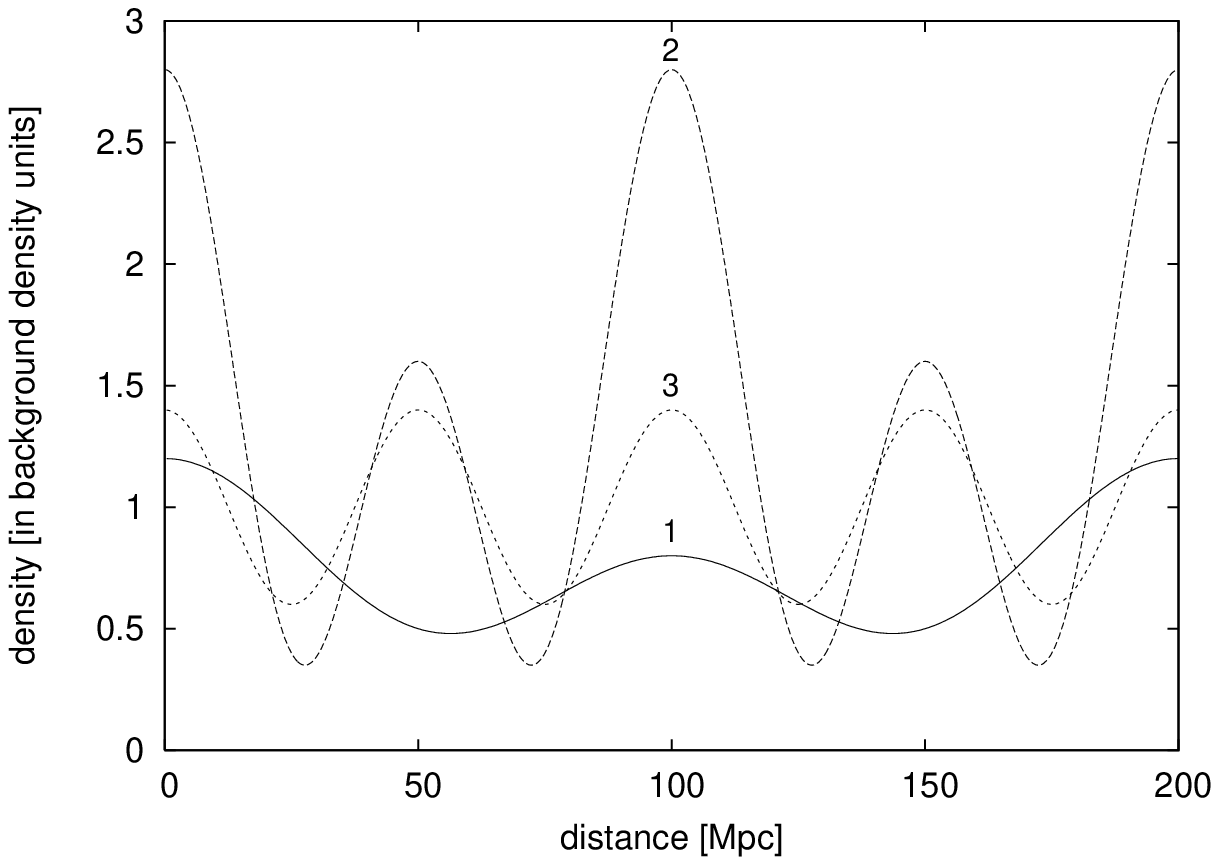}
\includegraphics[scale=0.4]{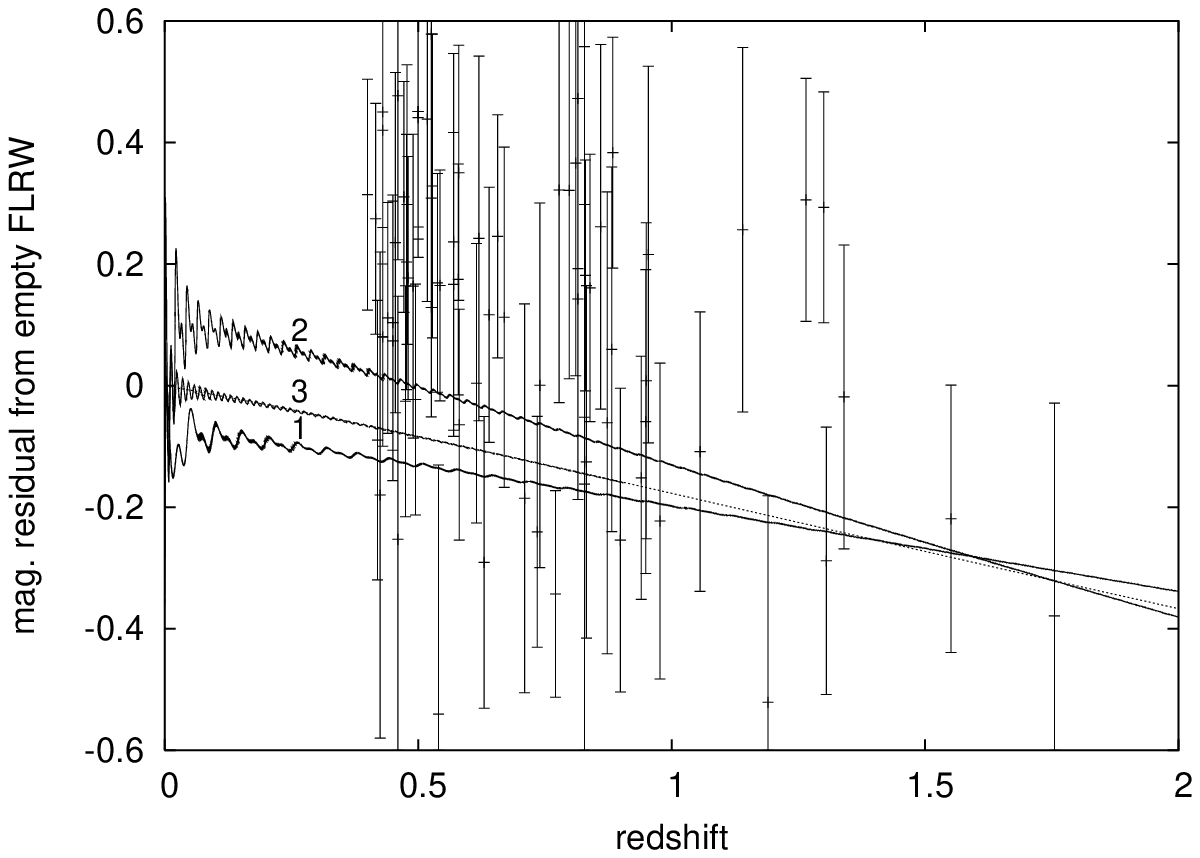} 
\caption{Density distribution (left panel) and the residual Hubble diagram
for models 1, 2 and 3.}
\label{f1}
\end{center}
\end{figure}

\begin{figure}
\begin{center}
\includegraphics[scale=0.4]{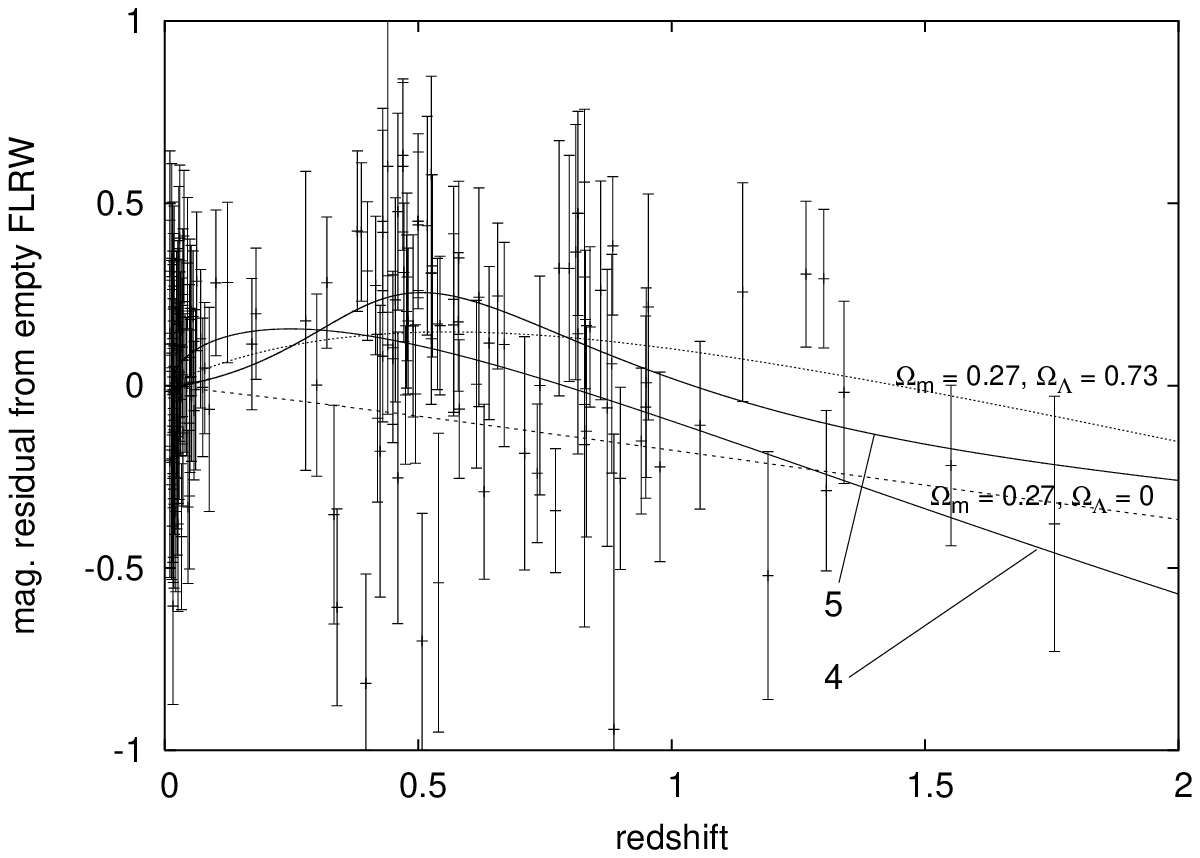}
\includegraphics[scale=0.4]{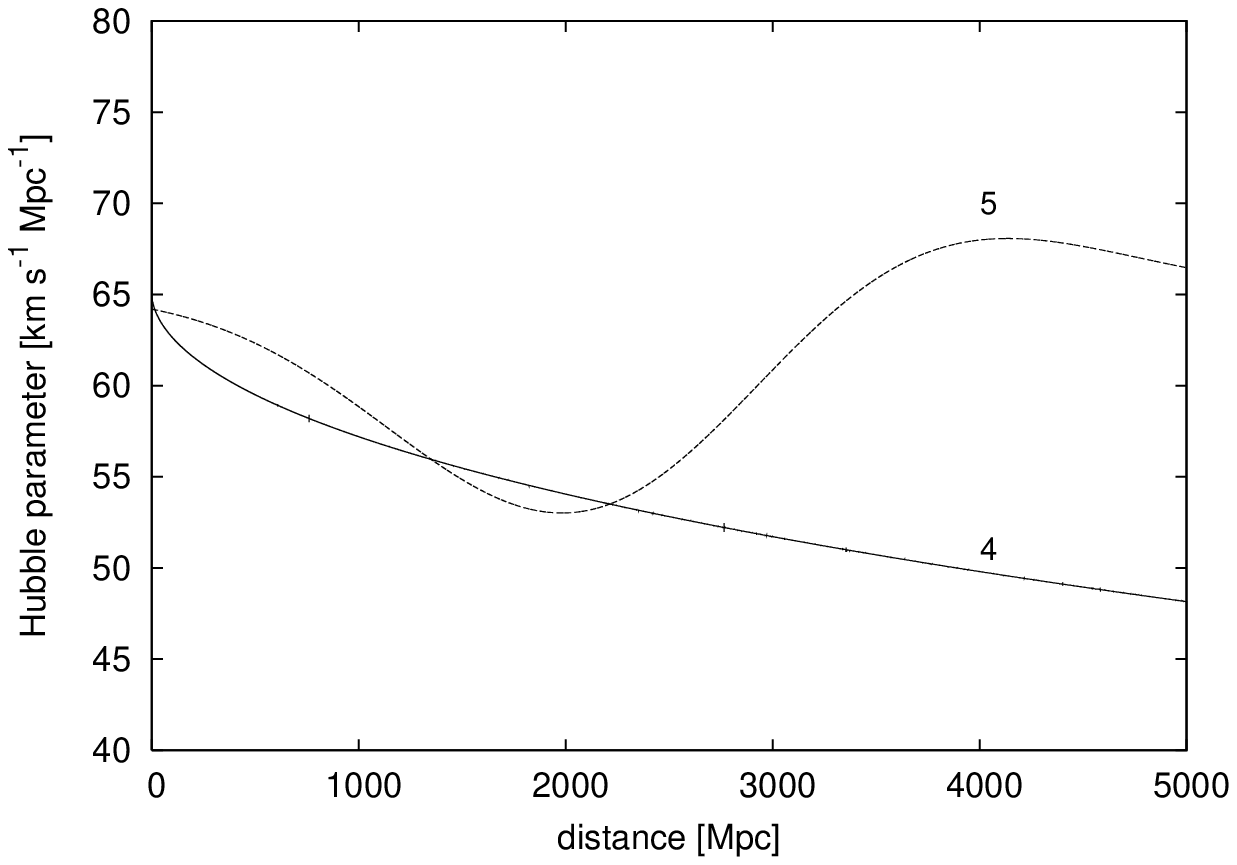} 
\caption{Residual Hubble diagram (left panel) and the Hubble parameter (right panel) for models 4 and 5.}
\label{f2}
\end{center}
\end{figure}

The supernova observations are analysed in the Lema\^itre--Tolman model.
Left panel of Fig. \ref{f1} presents three densities profiles used
to analyse the data. As one can see in right panel of Fig. \ref{f1}
these models do not fit supernova data well and  $\chi^2$/NDF 
is equal to 2.05, 1.46, 1.62 respectively for model 1, 2 and 3.
Models which fit supernova data quite well are presented in Fig. \ref{f2}.
The $\chi^2$/NDF for models 4 and 5 is 1.19, 1.15 respectively.
Density distribution in model 4 is increasing which suggest
that we are living near the centre of a large cosmic depression.
In model 5, the current density profile at present instant is homogeneous one.
However the evolution of model 5 is not as expected ---
it does not evolve from small initial fluctuations at last scattering instant.
In both of these models (models 4 and 5) expansion of the space is decreasing with
the distance. This feature can be seen in the right panel
of Fig. \ref{f2} where the Hubble parameter is presented.
The Hubble parameter is defined as a 1/3 of the scalar of the expansion, $H=(1/3) \Theta$.

\begin{figure}
\begin{center}
\includegraphics[scale=0.4]{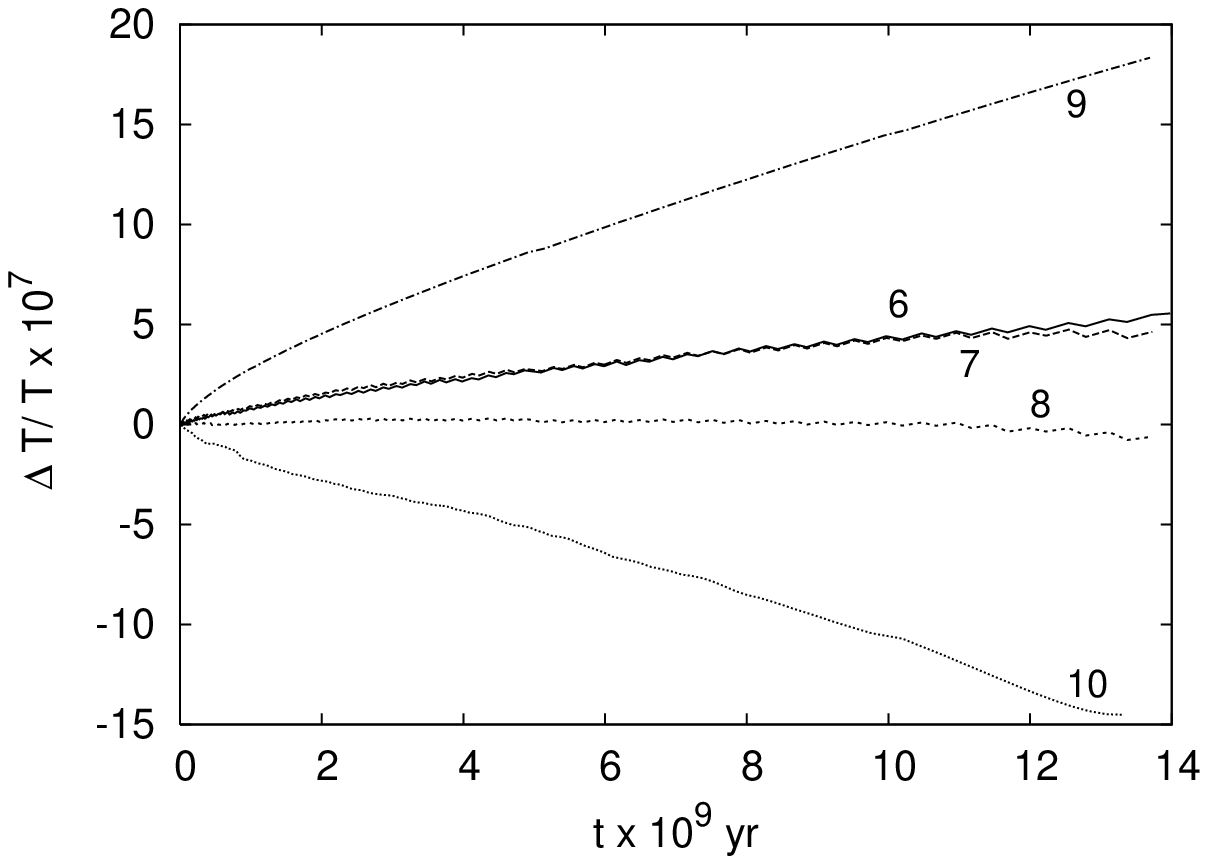} 
\includegraphics[scale=0.4]{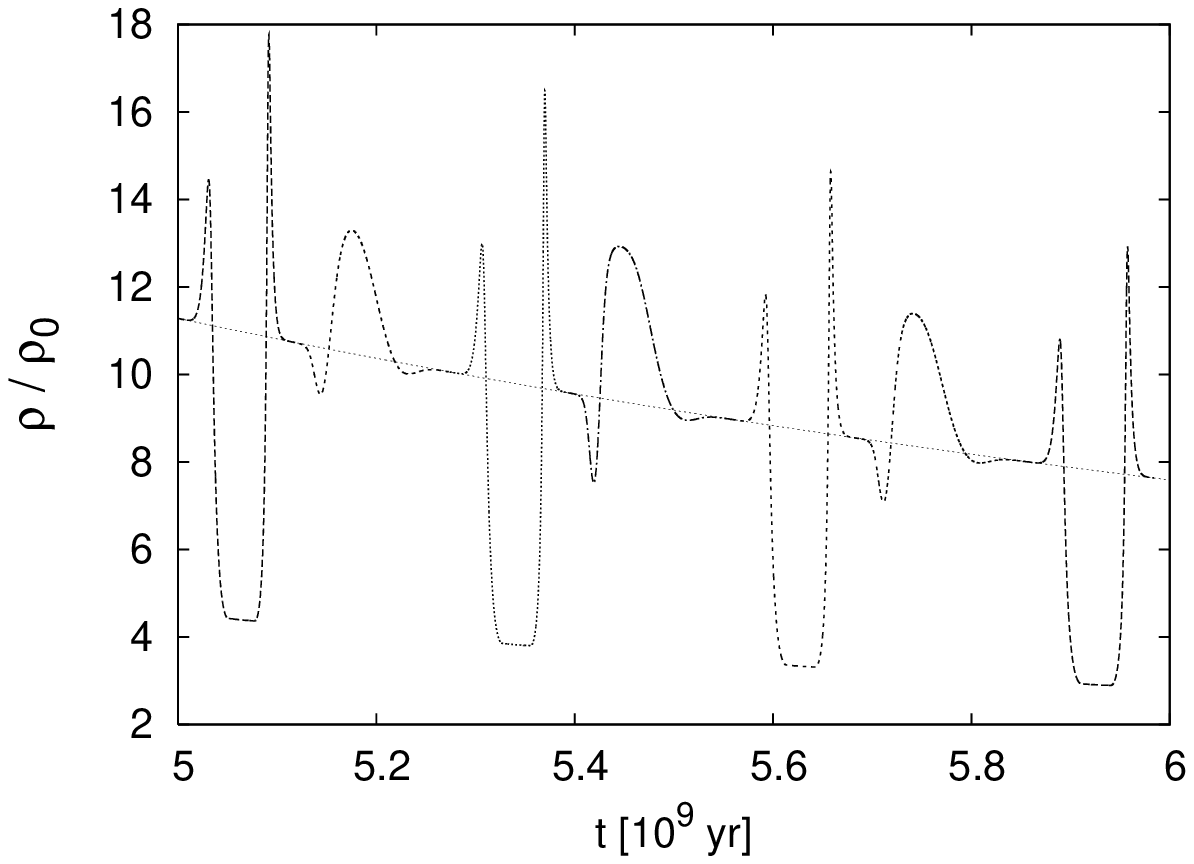}
\caption{Temperature fluctuations for models 6-10 (left panel). 
Fragment of density fluctuations in model 6 along which light propagates (right panel).}
\label{f3}
\end{center}
\end{figure}

The CMB observations are analysed in the Swiss cheese quasispherical Szekeres model.
Depending on the junction conditions, the final temperature fluctuations
caused just by the light propagation effects are of amplitude $10{-7}-10^{-6}$ 
which are one order of magnitude less than the observed temperature fluctuations.
The temperature fluctuations shown in the left panel of Fig. \ref{f3} are caused only by the light propagation effects and are evaluated for different moments of the Universe
evolution. Currently observed temperature fluctuations correspond to $t \approx 13.5 \times 10^9$y.
Right panel of Fig. \ref{f3} presents the fragment of density fluctuations
in model 6. 

\section{Conclusions}

The analysis of supernova observations was presented using the Lemaitre--Tolman model. It was shown that it is possible to fit these observations without the cosmological constant, but such models require that we live in a very special place in the Universe. On the other hand, density fluctuations of amplitude and size as observed in the local Universe are not able to fit observational data without the cosmological constant.
The CMB observations were analysed using the Szekeres model. It was shown that small-scale, non-linear inhomogeneities introduce temperature fluctuations of amplitude $10^{-6} - 10^{-7}$, which implies that small scale inhomogeneities do not influence the CMB data significantly via the Rees-Sciama effect. Thus, although matter distribution in the Universe is not homogeneous, and the impact of inhomogeneities on cosmological observations is visible, the results obtained using the inhomogeneous models do support the results obtained within the paradigm of concordance cosmology.

\section*{Acknowledgments}

The Copernicus Foundation for Polish Astronomy, the Warsaw Society of Sciences and the Polish Astroparticle Network (621/E-78/SN-0068/2007) are gratefully acknowledge for financial support.

\end{document}